# From Fully Supervised to Zero Shot Settings for Twitter Hashtag Recommendation

Abhay Kumar[1]*, Nishant Jain[1]*, Chirag Singh[1]♦, Suraj Tripathi[1]♦

[1] Samsung R&D Institute India - Bangalore
abhay1.kumar@samsung.com, nishant.jain@samsung.com, c.singh@samsung.com, suraj.tri@samsung.com

**Abstract.** We propose a comprehensive end-to-end pipeline for Twitter hashtags recommendation system including data collection, supervised training setting and zero shot training setting. In the supervised training setting, we have proposed and compared the performance of various deep learning architectures, namely Convolutional Neural Network (CNN), Recurrent Neural Network (RNN) and Transformer Network. However, it is not feasible to collect data for all possible hashtag labels and train a classifier model on them. To overcome this limitation, we propose a Zero Shot Learning (ZSL) paradigm for predicting unseen hashtag labels by learning the relationship between the semantic space of tweets and the embedding space of hashtag labels. We evaluated various state-of-the-art ZSL methods like Convex combination of Semantic Embedding (ConSE), Embarrassingly Simple Zero Shot Learning (ESZSL) and Deep Embedding Model for Zero Shot Learning (DEM-ZSL) for the hashtag recommendation task. We demonstrate the effectiveness and scalability of ZSL methods for the recommendation of unseen hashtags. To the best of our knowledge, this is the first quantitative evaluation of ZSL methods to date for unseen hashtags recommendations from tweet text.

**Keywords:** Zero Shot Learning, Deep Neural Network, Word Embedding, Twitter, Hashtag, Few Shot Learning.

## 1 Introduction

In recent years, Twitter has emerged as one of the most widely used microblogging and social networking website. Millions of active users produce a massive amount of tweets, targeted for information diffusion and social interaction. Tweets are microposts with a length constraint of 280 characters. These tweets span a wide range of topics, including political activities, personal posts, emerging social topics, and promotional content. Hashtags associated with tweets helps in categorization and retrieval of the posts based on the content and context. A Hashtag is a metadata tag in microblogs and consists of a string of characters prefixed with the "#" symbol. Hashtags have been used to organize tweets, facilitate easier search and propagate trendy topics

*, ♦ equal contribution

by creating an instant community with similar interests. Despite its effectiveness, only a small fraction of the tweets contains one or more hashtags. Hence, the motivation to automatically predict or recommend the hashtags has captured considerable attention of researchers recently. In addition, informal writing style and limited context due to the constraint in character length make it difficult to analyze tweets using traditional Natural Language Processing (NLP) models. However, the success of Deep Learning methods in diverse NLP tasks has accelerated research in social media analytics, including hashtag recommendation from tweets.

In this paper, we have proposed deep learning based hashtag recommendation methods for fully-supervised, zero-shot and few-shot settings. Additionally, we have presented the data collection and processing pipeline for Twitter hashtag dataset.

## 2 Related Work

Due to the increased adoption of hashtag usage and usefulness of hashtag recommendation, researchers have proposed multiple methods from various perspectives to predict predefined hashtags or general topic inherent in the tweet. Zangerle *et al.* [1] proposed hashtag recommendation system by retrieving tweets with similar content to that of the target tweet using *tf-idf* based content similarity and ranking the hashtags appearing in those tweets. Ding *et al.* [2] formulated the task as a topical translation model with the assumption that tweet content and hashtags represented the same theme but transliterated in different languages. Sedhai *et al.* [3] focused on hashtag recommendation for hyperlinked tweets based on retrieval of similar tweets and adopting RankSVM to rank the candidate hashtags. Inspired by [2], Gong *et al.* [4] proposed a non-parametric Bayesian method- Dirichlet Process Mixture Models with the similar assumption of parallel occurrences of tweet content and hashtags. These methods were based on the retrieval of related tweets from a large corpus and recommending hashtags based on content similarity and did not focus on the abstracted topics of tweets.

She *et al.* [5] developed a supervised topic model with hashtags as topic labels to learn the relationship among tweet words, hashtags, and topics. They assume that each tweet word comes as either background words (specific to the corpus) or local topic words (tweet specific) and generate hashtags using symmetric Dirichlet distribution. Zhang *et al.* [6] extended the translation-based models by considering temporal and personal factors. However, these traditional NLP techniques failed to capture the semantic and contextual information from a vast tweet corpus. Deep neural network based model architectures are capable of encoding rich contextual and semantic information and were experimented for hashtag recommendation task as well recently.

Weston *et al.* [7] proposed a CNN model for hashtag prediction considering it as a large-scale ranking task and represented both tweet words and entire tweet as embedding vectors. Gong *et al.* [8] suggested a deep CNN model with an attention mechanism to focus on the trigger words in tweets. Dey *et al.* [9] proposed two word-embedding based models, EmTaggeR, by learning word vector based on either global context or hashtag specific context.

Most of the research works have focused on hashtag recommendation in fully-supervised setting, where all hashtag labels are exposed in the training process. Only a few researchers have worked on unsupervised methods for the same. However, it is not feasible to collect data for all possible hashtags and train on them. This motivates for Zero Shot Learning based approaches to recommend unseen hashtags as well.

Zero Shot Learning (ZSL) methods have been experimented for various image classification task in recent years to expand classifiers to predict new class label even when there is no training sample for the same. Most of these methods involve training on annotated datasets of seen classes and modeling semantic relatedness of the unseen classes with respect to seen classes. Almost all earlier works on ZSL were targeted for the object classification task. Norouzi *et al.* [10] proposed a Convex Combination of Semantic Embedding (ConSE) model to map image features into semantic embedding space via convex combination of the class label embedding vectors. Romera-Paredes *et al.* [11] proposed a two linear layer network to learn the relationship between features, labels and attributes and model the network as a domain adaption method. Zhang *et al.* [12] proposed a ZSL model to use the features extracted from the Deep Neural Network (DNN) as the embedding space and combined multiple sematic modalities in an end-to-end learning fashion.

## 3 Dataset: Collection and Pre-Processing

In the following sub-sections, we have described the procedures to collect and prepare our Twitter hashtag dataset.

### 3.1 Data Collection

There is no available benchmark dataset for Twitter hashtag recommendation, because of the Twitter licensing conditions. Therefore, we collect the dataset using the Twitter API. We crawled 805K tweets during the last three weeks in January 2019.

- We collected tweets using Twitter *statuses/filter* stream API[1] with '#' as keyword.
- With the extension of tweet character length constraint to 280, Twitter API provides JSON field- *extended_tweet* to provide the complete text and metadata. For tweets identified by 'true' label for "truncated" field, we extracted *extended_tweet*.
- For tweets containing *retweeted_status* attribute, we recovered the original tweet by exploring the tweet embedded within the JSON field *retweeted_status*.

### 3.2 Data Cleaning and Preprocessing

We perform the following data cleaning steps to retain the useful tweets for our task.

- **Non-English tweets elimination:** We removed non-English tweets using the tweet language metadata provided with Twitter API.

[1] https://stream.twitter.com/1.1/statuses/filter.json

- **Non-hashtag tweets removal:** Since we wish to predict hashtag from tweets, we kept only those tweets, which have at least one hashtag and discarded other tweets.
- **Short-length tweets removal**: We deleted tweets containing less than five words, as it is challenging to get context out of concise tweets (e.g., what's on your mind).
- **Non-ASCII character removal & lower-case conversion:** We removed all non-ASCII character and converted all retained tweets into lowercase.
- **Normalization:** We replaced all user tags like '@xyz' with 'user'. Additionally, we also removed all web links from the tweets.
- **Duplicate tweets elimination:** We discarded duplicate tweets from the dataset.
- **Stop-words removal and tokenization:** We used NLTK[2] to remove the stop words. All tweets were tokenized using NLTK Twitter-aware *TweetTokenizer*.

### 3.3 Hashtag Extraction

We extracted and removed all the hashtags present in the tweets. There were 956 unique hashtags in the cleaned dataset. We kept the top 50 most used hashtags with each having at least 200 unique tweets to form a balanced dataset.

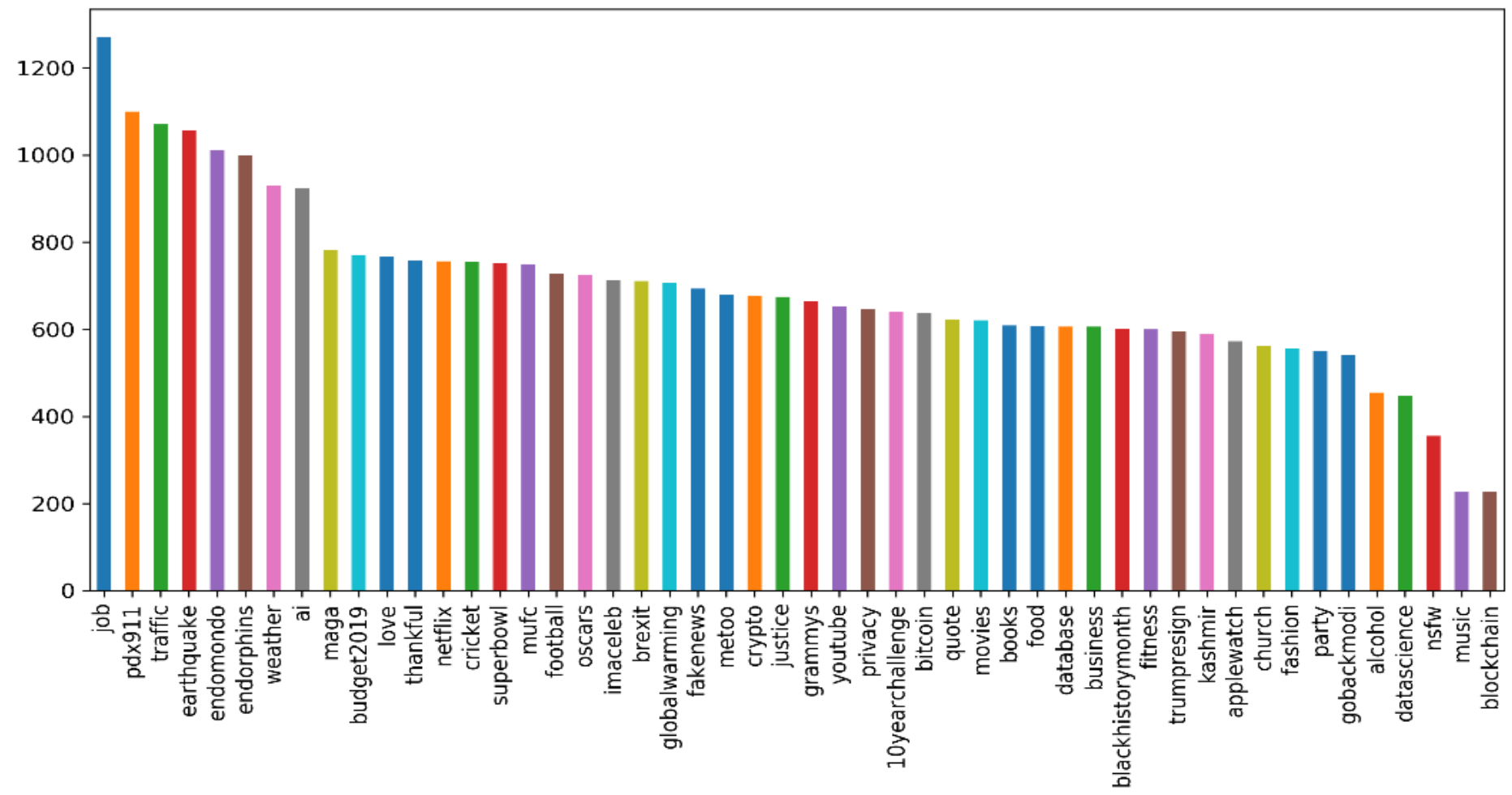

**Fig. 1.** Frequency distribution of tweets per hashtag.

### 3.4 Dataset Statistics and Visualization

We have shown (in Fig. 2) the word-cloud diagram to visualize the word frequencies in our Twitter hashtag dataset. Presence of similar font size words in the word-cloud shows that our dataset is balanced. We have also presented the t-SNE [13] plot of word2vec embedding vectors for the top 50 hashtags from our dataset. Hashtags frequently occurring together (like *crypto*, *bitcoin*, and *blockchain*) or semantically related (like *grammys* and *oscars*) are located nearby in the plot.

Table 1 shows some statistics for our data collection and cleaning steps.

[2] https://www.nltk.org/

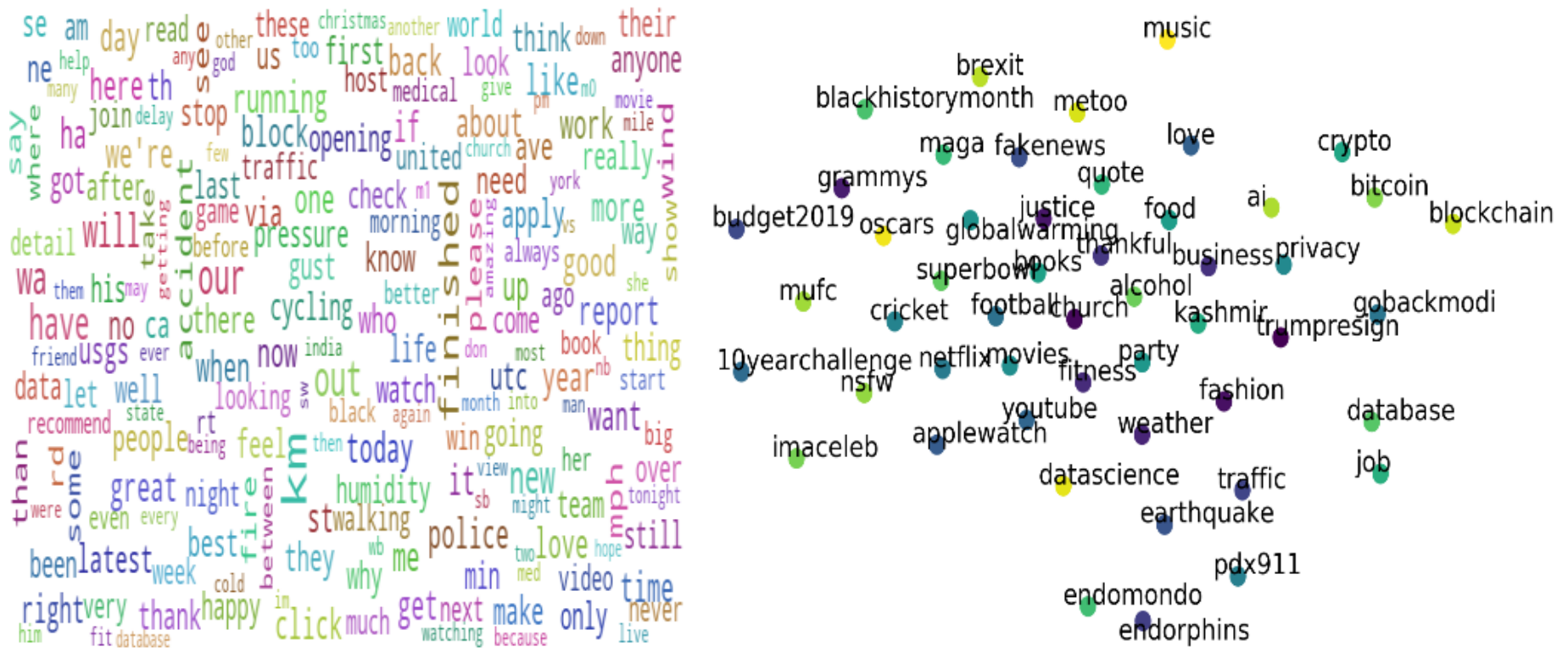

**Fig. 2.** Word-Cloud of tweet words and t-SNE visualization of hashtag embeddings.

**Table 1.** Twitter Hashtag Dataset Statistics

| Criteria | Count |
|---|---|
| Total Collected tweets | 805,031 |
| Cleaned sampled tweets | 321,385 |
| Tweets containing hashtags | 31,249 (9.7%) |
| Tweets containing more than one hashtag | 3,301 |
| Unique hashtags | 956 |
| Hashtags with at least 200 unique tweets | 50 |

## 4 Word Embedding for the Twitter dataset

Word embedding proposed by Mikolov *et al.* [14] is an unsupervised model to learn distributed representation of words. The use of word embedding has spread widely and rapidly in tasks like language modeling, text classification and sentiment analysis in recent years. The effectiveness of word embedding to extract the semantic and syntactic relationships between words has proved to be helpful in many NLP tasks.

Word embedding can be trained in two different configurations: - Continuous Bag of Words (CBOW) and skip-gram. In CBOW, the model predicts the current word based on a window of surrounding words while in skip-gram, the model tries to predict the surrounding context words based on the current word. According to the author's note[3] on the word2vec website, skip-gram provides better embedding for infrequent words while CBOW is faster for training. Thus, we use skip-gram to train our word embedding model.

Mikolov *et al.* [14] also provide a pre-trained word embedding trained on *Google News* dataset, which contains around 3 million words and phrases. However, the colloquial of tweets, which includes a large amount of informal language, abbreviations and slang, is quite different from the corpus of *Google News*. This leads to many out-of-vocabulary (OOV) words and thus gives poor performance.

---

3 https://code.google.com/archive/p/word2vec/

To overcome the above challenge, we trained our skip-gram model using 321,485 English tweets without using any preprocessing apart from the removal of URL links and replacing “@xyz” with “user”. The word embedding generated by our model has 94361 unique words, and each word is represented by a 150-dimensional vector. We have used this pre-trained embedding in all our experiments.

# 5 Proposed Models: Fully Supervised Settings

In the below subsections, we have presented the different model architectures. We have chosen the hyper-parameters for the same using the *sklearn library's GridSearchCV.* Categorical Cross-Entropy loss is used in all the models.

## 5.1 Artificial Neural Network (ANN)

ANN is a non-linear model, which maps the input embedding to output classes through hidden layers of fully connected neurons. To get the baseline, we use a single hidden layer of 1024 neurons with `tanh` as the activation function. In the output layer, Softmax layer is used to normalize the output of neurons into a probability distribution. A single neuron in hidden layer of ANN can be represented as:- $f_h(x) = \frac{e^x - e^{-x}}{e^x + e^{-x}}$. The output of final layer of ANN is calculated as $f_o(x_i) = \frac{e^{x_i}}{\sum_{j=1}^{c} e^{x_j}}$, where $i$ and $j$ goes from one to number of classes (c).

## 5.2 Convolutional Neural Network (CNN)

CNN comprises convolutional layers to extract feature based on the locality of reference in images. However, in recent years, their use has been extended to NLP domains as well. Kim [15] showed that CNN could be used effectively for text classification task also. We extract semantic feature map using convolution filter $f$ of shape $h \times d$, where $h$ is window length and $d$ is embedding dimension of the word. Our network employ 32 filters of window size 3, 5 and 7 in parallel on input embedding vectors. The generated feature maps by the three convolutional blocks are concatenated and fed to the hidden Fully Connected (FC) layer of 1024 and 256 neurons sequentially. The output of FC layer is fed to the softmax layer. Dropout is applied in both FC layers to counter the effect of overfitting.

## 5.3 Recurrent Neural Network (RNN)

Although CNN excels at extracting features from a local region but it fails to learn long-term dependencies. Therefore, RNN is used to learn characteristics from long-term dependencies through its recurrent structure. However, in practice, the vanilla RNN suffers from problems of short-term memory. It cannot retain relevant contextual information over longer sentences due to vanishing gradient problem. To overcome this problem, we use the Gated Recurrent Unit (GRU) [16] in our experiments, because of its ability to regulate the flow of information through it. We experiment with two different RNN architectures.

The first architecture consists of two GRUs. The first GRU unit outputs a vector of dimensionality 1024 for each state, which is fed to the second GRU. The output vector of the last state of second GRU is connected to a single FC layer with softmax activation function, which outputs the final probability density distribution of classes.

The second architecture consists of a single GRU of 1024 units along with the attention module. Attention module assigns a value between zero and one to each word depicting its relevance for the context. The final weighted output, obtained by multiplying each word's hidden state encoding vectors by its attention weights is fed to the final FC layer with softmax activation, which provides the class probabilities.

### 5.4 Transformer Network

Recently, the transformer network proposed by Vaswani *et al.* [17] has shown promising results by employing parallelization and unique use of multi-head attention module. The model has outperformed previous DNN architectures and has faster model training as compared to RNN (which is sequential by its design).

Transformer network takes the entire input sentence and adds positional encoding to each word vector. These word and positional embedding vectors are fed to the network, which consists of six identical attention modules. Each module comprises of a multi-head attention network and a feed-forward network. The network output is given directly to a FC layer with softmax activation, which predicts the hashtag probabilities. We use the same model architecture and hyper-parameters as in [17].

## 6 Proposed Models: Zero Shot Settings

**Zero Shot Learning:** In Zero-shot learning setting, we train a classifier from labeled training exemplars from seen classes and learn a mapping from input feature space to semantic embedding space. ZSL aims to classify class labels, which were never exposed during training pipeline. Let us assume that a training dataset, $D_{train} \equiv \{(x_i, y_i)\}_{i=1}^{m}$ is given, where $x_i \in \mathbb{R}^p$ represents the $p$-dimensional feature vector of each input tweet. Tweet semantic feature is 1024-dimensional encoding vector extracted from the pre-FC layer from RNN model (same as described in Section 5.3). There are $n_s$ distinct seen hashtag labels exposed during training such that $y_i \in Y_S \equiv \{1, 2, \ldots, n_s\}$. In testing dataset, $D_{test} \equiv \left\{(\acute{x}_j, \acute{y}_j)\right\}_{j=1}^{\acute{m}}$, where $\acute{x}_j \in \mathbb{R}^p$ and $\acute{y}_i \in Y_U \equiv \{n_s + 1, n_s + 2, \ldots, n_s + n_u\}$ represent tweet semantic feature and $n_u$ distinct unseen hashtag labels respectively. We have discussed the ZSL methods used for hashtag recommendation later in this section.

**Few Shot Learning:** Few Shot Learning (FSL) paradigm is based on feeding a limited amount of training data. This is an extension of ZSL setting, where few examples of unseen (in ZSL setting) class labels are also exposed during the training process. Therefore, the FSL model learns from training examples of all seen hashtag labels and few labeled examples of unseen hashtag labels. We have used the same ZSL methods and fed a few examples of unseen hashtags for our experiments.

### 6.1 ConSE: Convex Combination of Semantic Embeddings

ConSE [11] method does not learn the mapping $f: \mathcal{X} \rightarrow \mathcal{S}$ from the input space to semantic space by framing it as a regression model. Instead, it is formulated as a standard machine learning classifier, $p_\theta$ trained on $D_{train}$ to estimate the probability $p_\theta(y \mid x)$ of a tweet $x$ belonging to a seen hashtag label $y \in Y_S$. The ConSE method extends the prediction probability beyond the seen hashtags, to a set of unseen hashtag labels. The embedding vector of the unseen hashtag for a test tweet $x$ is predicted by a convex combination of seen hashtag embedding vectors weighted by their corresponding probabilities as shown in *Eq. 1.*

$$f(x) = \frac{1}{Z} \sum_{t=1}^{T} p_\theta(\hat{y}(x,t)|x).\, s(\hat{y}(x,t)) \tag{1}$$

Where $Z = \sum_{t=1}^{T} p_\theta(\hat{y}(x,t)|x)$ is the normalization factor, which represent the accumulated probability of top $T$ seen hashtags predicted for the tweet $x$ and $\hat{y}(x,t)$ denotes the $t^{th}$ most probable training hashtag label for a tweet $x$, and $s(\hat{y}(x,t))$ represents the embedding vector of $\hat{y}(x,t)$. Here, $T$ acts as a hyper-parameter and denotes the maximum number of seen hashtag embedding vectors to be considered. The value of $T$ is the same as the total number of seen hashtags in our experiments. Consequently, cosine similarity is used to find the most likely hashtag from unseen hashtag label as shown below-

$$\hat{y}(x,1) = \underset{\acute{y} \in Y_U}{\operatorname{argmax}} \cos(f(x), s(\acute{y})) \tag{2}$$

### 6.2 ESZSL: Embarrassingly Simple Approach to Zero Shot Learning

ESZSL [12] approach is a general framework which models the relationship among features, attributes, and class labels by formulating it as a two linear layers network. In training stage, let $\mathcal{A} \in \mathbb{R}^{s \times n_s}$ denote the attribute space with $s$ attributes per seen hashtag label, $\mathcal{X} \in \mathbb{R}^{p \times m}$ and $\mathcal{Y} \in \{0,1\}^{m \times n_s}$ represent the tweet semantic features and hashtag embedding vectors respectively for all $m$ training examples. Tweet feature vector is 1024-dimensional vector extracted from the pre-FC layer from RNN model (same as described in Section 5.3). Each row of $\mathcal{Y}$ is one-hot label with one corresponding to the true hashtag label and zero corresponding to the rest of the hashtags. We minimize the following loss function to learn a linear predictor $\mathcal{W}$ from all $m$ training examples.

$$\underset{\mathcal{W} \in \mathbb{R}^{p \times s}}{\text{minimize}} \ \mathcal{L}(\mathcal{X}^T \mathcal{W} \mathcal{A},\ \mathcal{Y}) + \Phi(\mathcal{W}) \tag{3}$$

where $\mathcal{W}$ represents the parameters to be learned, and $\Phi$ is the *Frobenius norm* regularization. At inference time, we predict the hashtag label by $argmax_i(\mathrm{x}^T \mathcal{W} \hat{\mathcal{A}}_i)$, where $\hat{\mathcal{A}} \in \mathbb{R}^{s \times n_u}$ contains $s$-dimensional attribute vector for all $n_u$ unseen hashtag labels. The closed form solution for *Eq.* 3 is expressed as-

$$\mathcal{W} = (\mathcal{X}\mathcal{X}^T + \gamma I)^{-1} \mathcal{X}\mathcal{Y}\mathcal{A}^T (\mathcal{A}\mathcal{A}^T + \gamma I)^{-1} \tag{4}$$

We have used trained word2vec 150-dimensional embedding vector as the attribute vector ($s$ = 150) for each hashtag label for our experimentations.

### 6.3 DEM-ZSL: Deep Embedding Model for Zero Shot Learning

DEM-ZSL [13] is end-to-end learning of deep embedding model with two branches. One branch learns the semantic space of the tweet and has the same architecture as the RNN (mentioned in Section 5.3) without the FC and softmax layers. The other branch learns the semantic representation of the hashtag class labels. We use the trained word2vec to get the embedding of hashtag labels. These embedding vectors of hashtag labels are fed to a FC layer (with 1024 neurons) with Rectified Linear Unit (ReLU) activation. This maps the hashtag embedding vectors of 150 dimensions into 1024-dimensional semantic space, which represents different combinations of the embedding features. Finally, we minimize the least square error to reduce the discrepancy between tweet semantic features and the mapped 1024-dimensional semantic vector of hashtag labels.

## 7 Evaluations and Discussions

We have performed multiple experiments for Twitter hashtag recommendation ranging from fully supervised setting to zero-shot setting. The experimental details and results for the same are discussed in the following sub-sections. For all our experiments, we use *Xavier uniform initializer (bias=0)* for initializing all layers and *Adam (learning rate = 0.001)* optimizer for back-propagating the error.

### 7.1 Evaluation matrix

In the supervised setting, we present the following metrics for evaluation:-

$$Accuracy = \frac{Correct\ Hashtag\ Predictions}{Total\ number\ of\ tweets};\ Precision = \frac{TP}{TP+FP} \quad (5)$$

$$Recall = \frac{TP}{TP+FN};\ F1\ score = 2 * \frac{Precision*Recall}{Precision+Recall} \quad (6)$$

where TP is True Positive, FP is False Positive and FN is False Negative.

In the unsupervised setting, we present the accuracy in terms of *Flat-Hit@K,* which represents the percentage of test tweet examples for which model predicts the true hashtag label among the top K predicted hashtags.

### 7.2 Evaluation of fully supervised setting

We have experimented with different supervised-learning model architectures for Twitter hashtag recommendation. For the supervised setting, we exposed tweets corresponding to all hashtags in the training phase. We have used stratified k-fold 80-20 split and presented the 5-fold cross validation accuracies for all models in Table 2. The Transformer Network, with a shorter path length of $O(1)$ outperforms CNN and RNN models with $O(\log n)$ and $O(n)$ path lengths respectively for the forward and backward signals to traverse. Shorter path length in Transformer Network helps in learning long-range dependencies better [18].

**Table 2.** Experimental results for hashtag recommendation in fully supervised setting

| Models | Accuracy (%) | Precision | Recall | f1 score |
|---|---|---|---|---|
| ANN (Baseline) | 40.7 | 0.41 | 0.41 | 0.41 |
| CNN | 43.7 | 0.44 | 0.44 | 0.44 |
| RNN | 46.9 | 0.47 | 0.47 | 0.47 |
| RNN with Attention | 47.0 | 0.47 | 0.47 | 0.47 |
| Transformer Network | **57.4** | **0.57** | **0.57** | **0.57** |

### 7.3 Evaluation of zero-shot and few-shot setting

Additionally, we have also experimented with different Zero-Shot Learning (ZSL) and Few-Shot Learning (FSL) model architectures for Twitter hashtag recommendation from unseen hashtags. In the ZSL setting, we divide our hashtag class labels into seen and unseen categories and train our models on seen hashtags only. Similarly, in FSL settings, we train our model with a few (random count ranging from 5-10) tweet examples from the unseen hashtag classes as well. We present the flat hit accuracies for top 1, 2, and 5 hashtags prediction accuracy for all model architectures in Table 3. Mapping the hashtag word2vec embedding vector to an intermediate embedding space of 1024 helps to capture the various combination of the original embedding features. Hence, DEM-ZSL method achieves better hashtag recommendation *Flat-Hit* percentage and better understands the relation between the semantic space of tweets and the embedding space of hashtag labels by addressing the *hubness* problem [12].

**Table 3.** Experimental results for hashtag recommendation in ZSL and FSL settings

| Seen/Unseen Hashtags | ZSL Model | Zero Shot Setting (in percentage) | | | Few Shot Setting (in percentage) | | |
|---|---|---|---|---|---|---|---|
| | | **hit@1** | **hit@2** | **hit@5** | **hit@1** | **hit@2** | **hit@5** |
| 40/10 | ConSE | 49 | 65 | 88 | 54 | 66 | 89 |
| | ESZSL | 60 | 73 | 93 | 64 | 75 | 93 |
| | DEM | **62** | **74** | **96** | **73** | **83** | **97** |
| 30/20 | ConSE | 27 | 36 | 59 | 43 | 50 | 67 |
| | ESZSL | 35 | 48 | **71** | 46 | 56 | 73 |
| | DEM | **42** | **54** | **71** | **61** | **71** | **84** |
| 25/25 | ConSE | 24 | 30 | 45 | 38 | 47 | 62 |
| | ESZSL | 23 | 35 | **61** | 40 | 49 | 66 |
| | DEM | **29** | **42** | 59 | **58** | **69** | **82** |

### 7.4 Discussions

**Effect of different word embedding initializations:** We have tried out both pre-trained on twitter dataset and google word2vec embedding in our models. We observe a significant improvement in using pre-trained embedding. In addition, we experimented with different dimensions (150 and 300) of word embedding vectors, i.e. and achieved very similar accuracies. Therefore, we decide to use 150-dimensional embedding to save computation cost without affecting the accuracies.

**Hyper-parameter tuning:** We have performed a number of experiments heuristically using grid search methodology to obtain the optimal hyper-parameters for the different models. We have experimented with different activation functions and decided to use *SoftReLU* for convolutional layers and *ReLU* for GRU and FC layers respectively.

**Accuracy Discussion:** In the supervised setting, transformer network can capture long-range discourse structure from the tweet in order to get the complete context for recommending the correct hashtag. For ZSL and FSL settings, end-to-end optimization in DEM-ZSL leads to a better intermediate embedding space of hashtags for improved ZSL performance. Three different splits (40/10, 30/20, 25/25) of seen and unseen hashtags have been tried to evaluate the scalability of the ZSL models. Even with 25/25 split, ZSL method recommends relevant hashtags with around 60% hit@5.

**Hashtag recommendation in the wild:** The use of hashtags in Twitter is very dynamic and frequently changes. Since supervised models can recommend hashtags from the seen hashtag labels only, the model needs to be re-trained to accommodate any newly created hashtag. Therefore, we proposed ZSL and FSL methods to predict unseen hashtags, although trained on seen hashtags only. We can use external prior (like trending hashtags, all-time famous hashtags) to decide the unseen hashtags for a given scenario and employ the ZSL methods to predict the most probable hashtags from the set of unseen hashtags. In Table 4, we have shown few representative tweets and their hashtag recommendations.

**Table 4.** Hashtag recommendation results for few tweet examples

| Cleaned Tweets | Expected Hashtag | Top 5 Hashtags |
|---|---|---|
| high time girls women troubled under user should come out on there are cases of girls exploited need to give courage to those souls to take up on him | #metoo | **#metoo**, #justice, #netflix, #privacy, #fakenews |
| fact: The uncertainty of a parameter estimate goes to zero as the sample size approaches infinity. The variability of a parameter estimate does not. | #datascience | #ai, **#datascience**, #church, #privacy, #bitcoin |
| So it's obvious the refs don't want us to win, but guess what bitches, here we come champs | #superbowl | **#superbowl**, #church, #fitness, #movies, #thankful |
| History of gun violence is long & still unresolved & unsolved for many victims & those seeking | #justice | #trumpresign, **#justice**, #fakenews, #metoo, #church |
| wtf why should we fill up their pockets, anybody who is paying to watch these 2 is a direct victim of msm | #fakenews | **#fakenews**, #movies, #metoo, #privacy, #trumpresign |

# 8 Conclusions

In the paper, we have experimented with various deep learning based model for Twitter hashtag recommendation. We have presented the experimental result both in the supervised and zero-shot setting. The ZSL models can predict unseen hashtags, even if those hashtags are not exposed in the training phase. These ZSL models learn the mapping from semantic space of tweets to hashtags embedding space. For determining the semantic space of tweets, we have implemented CNN and RNN based encoders. Additionally, we have demonstrated the statistical analysis of the dataset collected for this work. For future work, we can also try out multimodal inputs, like tweets text and images and learn a common visual-text embedding space and learn mapping to the embedding space of hashtag labels for hashtag recommendation.